\documentclass[journal]{IEEEtran}
\usepackage[bookmarks,colorlinks]{hyperref}
\usepackage{enumitem}
\usepackage[linesnumbered,ruled,lined]{algorithm2e}

\usepackage{graphicx}
\usepackage{dcolumn}
\usepackage{bm}

\usepackage[usenames,dvipsnames]{xcolor}

\usepackage{tabularx}
\usepackage{array}

\usepackage{amsmath,amsfonts}
\usepackage{mathtools} 

\usepackage{array}
\usepackage[caption=false,font=normalsize,labelfont=sf,textfont=sf]{subfig}
\usepackage{textcomp}
\usepackage{stfloats}
\usepackage{url}
\usepackage{verbatim}
\usepackage{graphicx}
\usepackage{cite}
\usepackage{svg}
\usepackage[noend]{algpseudocode} 
\usepackage{etoolbox}
\usepackage{booktabs}   
\usepackage{array}      
\usepackage{multirow}

\usepackage{float}
\usepackage{pgfplots}
\usepackage[bookmarks,colorlinks]{hyperref}
\usepackage[linesnumbered,ruled,lined]{algorithm2e}
\usepackage{enumitem}
\usepackage{algpseudocode}
\usetikzlibrary{shapes.multipart,intersections}
\usepackage{cite}
\usepackage{amsmath,amssymb,amsfonts,amsthm,steinmetz}
\usepackage{graphicx}
\usepackage{mathrsfs}  
\usepackage{textcomp}
\usepackage{acronym}
\usepackage{xcolor}
\usepackage{upgreek,xspace}

\usepackage{tikz}
\usetikzlibrary{calc}
\makeatletter
\newcommand{\gettikzxy}[3]{%
  \tikz@scan@one@point\pgfutil@firstofone#1\relax
  \edef#2{\the\pgf@x}%
  \edef#3{\the\pgf@y}%
}
\makeatother

\usepackage[draft]{todonotes}

\usepackage{esvect}

\usepackage{times}
\usepackage{bm}
\usepackage{amsmath}
\usepackage{amssymb}
\usepackage{stmaryrd}
\usepackage{babel}
\usepackage{graphics, graphicx}
\usepackage{xcolor}
\usepackage{gensymb}
\usepackage{cite}
\usepackage{enumitem}
\usepackage{url}

\ifCLASSINFOpdf
\else
\fi
\begin{document}
%
\title{Virtual VNA 3.1: Non-Coherent-Detection-Based Non-Reciprocal Scattering Matrix Estimation Leveraging a Tunable Load Network}
%
%
%

\author{Philipp~del~Hougne,~\IEEEmembership{Member,~IEEE}
\thanks{This work was supported in part by the ANR France 2030 program (project ANR-22-PEFT-0005), the ANR PRCI program (project ANR-22-CE93-0010), the European Union's European Regional Development Fund, and the French region of Brittany and Rennes Métropole through the contrats de plan État-Région program (projects ``SOPHIE/STIC \& Ondes'' and ``CyMoCoD'').}
\thanks{P.~del~Hougne is with Univ Rennes, CNRS, IETR - UMR 6164, F-35000 Rennes, France (e-mail: philipp.del-hougne@univ-rennes.fr).}
}

\maketitle

\begin{abstract}
We refine the recently introduced ``Virtual VNA 3.0'' technique to remove the need for coherent detection. The resulting ``Virtual VNA 3.1'' technique can unambiguously estimate the full scattering matrix of a non-reciprocal, linear, passive, time-invariant device under test (DUT) with $N$ monomodal ports using an $N_\mathrm{A}$-channel coherent wavefront generator and an $N_\mathrm{A}$-channel \textit{non}-coherent detector, where $N_\mathrm{A}<N$. Waves are injected and received only via a fixed set of $N_\mathrm{A}$ ``accessible'' DUT ports while the remaining $N_\mathrm{S}$ ``not-directly-accessible'' DUT ports are terminated by a specific tunable load network. To resolve all ambiguities, an additional modified setup is required in which waves are injected and received via a known $2N_\mathrm{A}$-port system connected to the DUT's accessible ports.
We experimentally validate our method for $N_\mathrm{A}=N_\mathrm{S}=4$ considering a non-reciprocal eight-port circuit as DUT.
By eliminating the need for coherent detection, our work reduces the hardware complexity which may facilitate applications to large-scale or higher-frequency systems. Additionally, our work provides fundamental insights into the minimal requirements to fully and unambiguously characterize a non-reciprocal DUT.
\end{abstract}


\IEEEpeerreviewmaketitle

\section{Introduction}
\label{sec_introduction}

A linear, passive, time-invariant device under test (DUT) with $N$ monomodal lumped ports is fully characterized by its scattering matrix $\mathbf{S}\in\mathbb{C}^{N \times N}$. At microwave frequencies, $\mathbf{S}$ is conventionally measured by connecting each port of the DUT via a monomodal transmission line to a distinct port of an $N$-port vector network analyzer (VNA). The VNA then injects and receives waves via the $N$ DUT ports to determine $\mathbf{S}$. In certain scenarios, however, a portion of the DUT ports might be not directly accessible (NDA), precluding the injection and reception of waves via these NDA ports. Consider, for instance, a scenario in which the $N$-port DUT has more ports than the available $N_\mathrm{A}$-port VNA (i.e., $N>N_\mathrm{A}$). In principle, one may then determine $\mathbf{S}$ by reconnecting the VNA multiple times to different DUT ports while terminating the other DUT ports with matched loads~\cite{tippet1982rigorous,ruttan2008multiport,2023paper}. However, such procedures require repeated manual reconnections which are error-prone and time-consuming, and therefore not scalable to scenarios with $N \gg N_\mathrm{A}$. Instead, it would be appealing to only connect the VNA once to a subset of $N_\mathrm{A}$ DUT ports and to never inject or receive waves via the DUT's remaining $N_\mathrm{S}=N-N_\mathrm{A}$ NDA ports. 

The recently introduced ``Virtual VNA'' technique~\cite{del2024virtual,del2024virtual2p0,del2025virtual3p0,kitvna} enables exactly that. Specifically, provided that the DUT's NDA ports are terminated by a suitable tunable load network\footnote{The tunable load network must switch the termination of each of the DUT's NDA ports between three distinct and known individual loads as well as known coupled loads connecting it to neighboring DUT ports~\cite{del2024virtual2p0,del2025virtual3p0}.}, it is indeed possible to retrieve $\mathbf{S}$ fully and free of ambiguities by measuring the $N_\mathrm{A}\times N_\mathrm{A}$ scattering matrix at the DUT's accessible ports for a set of different terminations of the DUT's NDA ports~\cite{del2024virtual,del2024virtual2p0,del2025virtual3p0,kitvna}. Each port of the required tunable load network can then be interpreted as an additional ``virtual'' VNA port, such that the combination of the $N_\mathrm{A}$-port VNA with the $N_\mathrm{S}$-port tunable load network yields a ``Virtual VNA'' with $N=N_\mathrm{A}+N_\mathrm{S}$ ports.
Detailed comparisons of the ``Virtual VNA'' technique to related literature~\cite{garbacz1964determination,bauer1974embedding,mayhan1994technique,davidovitz1995reconstruction,wiesbeck1998wide,lu2000port,lu2003multiport,pfeiffer2005characterization,pfeiffer2005recursive,pfeiffer2005equivalent,pursula2008backscattering,bories2010small,denicke2012application,monsalve2013multiport,van2020verification,sahin2021noncontact,buck2022measuring,kruglov2023contactless,sol2024experimentally,sol2024optimal,shilinkov2024antenna,del2025physics} can be found in~\cite{del2024virtual,del2024virtual2p0,del2025virtual3p0} and are not repeated here for brevity.

So far, our discussion treats the $N_\mathrm{A}$-port VNA as a black box that determines the $N_\mathrm{A} \times N_\mathrm{A}$ measurable scattering matrix at the DUT's accessible ports for each considered load configuration terminating the DUT's NDA ports. In fact, besides a conventional VNA, any $N_\mathrm{A}$-port apparatus capable of injecting and receiving coherent wavefronts could be used to measure the $N_\mathrm{A} \times N_\mathrm{A}$ measurable scattering matrix at the DUT's accessible ports. In general, the larger $N_\mathrm{A}$ is, the better the ``Virtual VNA'' technique can cope with measurement noise. This consideration may motivate the assembly of a custom-built apparatus because many-port VNAs are extremely costly. The generation of an $N_\mathrm{A}$-channel coherent wavefront can be realized with a single continuous-wave source, an $N_\mathrm{A}$-way power divider and $N_\mathrm{A}$ IQ modulators. The separation of waves that are ingoing and outgoing via the DUT's ports can be realized with circulators. But $N_\mathrm{A}$-channel coherent reception may be a bottleneck due to its requirement for synchronization. In contrast, $N_\mathrm{A}$-channel \textit{non-}coherent detection is drastically simpler to realize, for instance, with a single‐pole $N_\mathrm{A}$‐throw switch and a single-port spectrum analyzer. More generally, reducing hardware complexity is a key enabler for large-scale or higher-frequency systems. These considerations motivate the investigation of whether the ``Virtual VNA'' technique can cope with a constraint to non-coherent detection. In addition to this technological motivation, understanding the minimal requirements for estimating $\mathbf{S}$ is a curiosity of fundamental interest.

At its core, the raised question relates to the ability to retrieve the phases of some transfer matrix $\mathbf{H}\in\mathbb{C}^{P_2\times P_1}$ based on measurements of $|\mathbf{y}|^2 = |\mathbf{H}\mathbf{x}|^2$, where $\mathbf{x}\in\mathbb{C}^{P_1}$ and $\mathbf{y}\in\mathbb{C}^{P_2}$ are the $P_1$-channel input wavefront and the $P_2$-channel output wavefront, respectively. Indeed, if a method exists to unambiguously retrieve $\mathbf{H}$ based on injecting a set of known $\mathbf{x}$ and detecting the corresponding set of $|\mathbf{y}|^2$, then this method can be used to replace the VNA in the ``Virtual VNA'' technique (with $\mathbf{H}$ being the measurable scattering matrix at the DUT's accessible ports). As such, we are facing a standard phase-retrieval problem~\cite{shechtman2015phase,dong2023phase}. In general, based on a sufficiently large and diverse set of pairs $\{\mathbf{x},|\mathbf{y}|^2\}$, it is fairly easy to retrieve an estimate of $\mathbf{H}$~\cite{dremeau2015reference,metzler2017coherent,caramazza2019transmission} up to row-wise phase ambiguities~\cite{dremeau2015reference,del2016intensity,goel2023referenceless}. These row-wise phase ambiguities fundamentally arise from 
\begin{equation}
    \left|y_i\right|^2 = \left|\sum_{j=1}^{P_1}  H_{ij} x_j\right|^2 = \left|\sum_{j=1}^{P_1} \left( \mathrm{e}^{\jmath \theta_i} H_{ij} \right)x_j\right|^2 \ \forall \ \theta_i,
    \label{eq_1}
\end{equation}
where $x_i$, $y_j$ and $H_{ij}$ denote, respectively, the $i$th entry of $\mathbf{x}$, the $j$th entry of $\mathbf{y}$, and the $(i,j)$th entry of $\mathbf{H}$. In the special case of $\mathbf{H}$ being a symmetric square matrix, the row-wise phase ambiguities collapse to a global phase ambiguity, i.e., $\left|\mathbf{y}\right|^2 = \left| \mathrm{e}^{\jmath\theta}\mathbf{H}\mathbf{x}\right|^2 \ \forall \ \theta$, because the symmetry constraint ties the row phases and column phases together. A global phase ambiguity does not adversely affect most conceivable applications. Row-wise phase ambiguities, however, can seriously hinder applications in coherent wave control (e.g., beamforming).

Three important classes of $\mathbf{H}$ exist in wave engineering contexts. 
\textit{First}, $\mathbf{H}$ can be an off-diagonal block of a scattering matrix, implying that it has no symmetry and is not necessarily square. Relevant examples arise in optical wavefront shaping (``transmission matrix'') and wireless MIMO communications (``channel matrix''). 
\textit{Second},  $\mathbf{H}$ can be a reciprocal scattering matrix which is square and symmetric. 
\textit{Third}, $\mathbf{H}$ can be a non-reciprocal scattering matrix which is square but not symmetric. Only in the second case there is no issue with row-wise phase ambiguities. Accordingly, for reciprocal DUTs, we previously showed the compatibility of the corresponding ``Virtual VNA 2.0'' gradient-descent technique with non-coherent detection in~\cite{del2024virtual2p0}.\footnote{Prior to~\cite{del2024virtual2p0}, we already explored the limitation to non-coherent detection in characterizing reciprocal DUTs in earlier works~\cite{sol2024experimentally,del2024virtual} that were restricted to less elaborate tunable loads, preventing the lifting of all ambiguities in $\mathbf{S}$.}

The row-wise phase ambiguities in the first and third case can be lifted based on the non-coherent detection of known interferences of the outgoing wavefronts. In an optical wavefront shaping context,~\cite{goel2023referenceless} lifted the row-wise phase ambiguities of a transmission matrix by modulating the outgoing wavefront with known random configurations of a spatial light modulator and measuring the resulting intensities. The method proposed in~\cite{goel2023referenceless} relied on an assumption of forward-only scattering which substantially simplifies the mathematical description but cannot be used in our problem of characterizing a non-reciprocal DUT with lumped ports. 

Our contributions in this paper are as follows. \textit{First}, we demonstrate how to lift the row-wise phase ambiguities arising when retrieving the phases of an $N_\mathrm{A}$-port non-reciprocal scattering matrix given an $N_\mathrm{A}$-channel coherent wavefront generator and an $N_\mathrm{A}$-channel \textit{non-}coherent wavefront detector. Specifically, besides non-coherent measurements on the $N_\mathrm{A}$-port system, we also perform non-coherent measurements on the cascade of this system with a known $2N_\mathrm{A}$-port system. We rigorously account for multiple scattering between the two systems via a multi-port network formulation. \textit{Second}, we apply this method to upgrade the ``Virtual VNA 3.0'' gradient-descent technique so that we can unambiguously estimate the full $N\times N$ scattering matrix of a non-reciprocal DUT by connecting $N_\mathrm{S}$ DUT ports to a specific tunable load network, exciting the $N_\mathrm{A}$ remaining DUT ports with known coherent wavefronts and only measuring the intensities of the outgoing wavefronts with a non-coherent detector.
\textit{Third}, we validate the resulting ``Virtual VNA 3.1'' technique experimentally, considering a non-reciprocal eight-port circuit (a complex transmission-line network
containing circulators) with $N_\mathrm{A}=N_\mathrm{S}=4$. We examine origins of inaccuracies through careful benchmarking.

The remainder of this paper is organized as follows. 
In Sec.~\ref{sec_problem_statement}, we define the measurement problem studied in this paper. 
In Sec.~\ref{sec_methods}, we present a general method for ambiguity-free phase retrieval (Sec.~\ref{subsec_PR}) and then leverage it to propose the ``Virtual VNA 3.1'' technique (Sec.~\ref{V2NA3p1}).
In Sec.~\ref{sec_ExpVal}, we experimentally validate the ``Virtual VNA 3.1'' technique.
We close with a conclusion and outlook in Sec.~\ref{sec_conclusion}.

\section{Problem Statement}\label{sec_problem_statement}

Our goal is to unambiguously estimate the full scattering matrix $\mathbf{S}\in\mathbb{C}^{N\times N}$ of a linear, passive, time-invariant and non-reciprocal DUT with $N$ monomodal lumped ports. The DUT's non-reciprocity implies that $\mathbf{S}$ is not symmetric: $\mathbf{S}\neq\mathbf{S}^\top$. No a priori knowledge about the DUT is available. The DUT's ports are partitioned into two fixed sets: the set $\mathcal{A}$ includes the indices of accessible DUT ports via which waves may be injected and received; the set $\mathcal{S}$ includes the indices of the remaining NDA DUT ports which are terminated by a specific tunable load network. 

\begin{figure}
    \centering
    \includegraphics[width=\columnwidth]{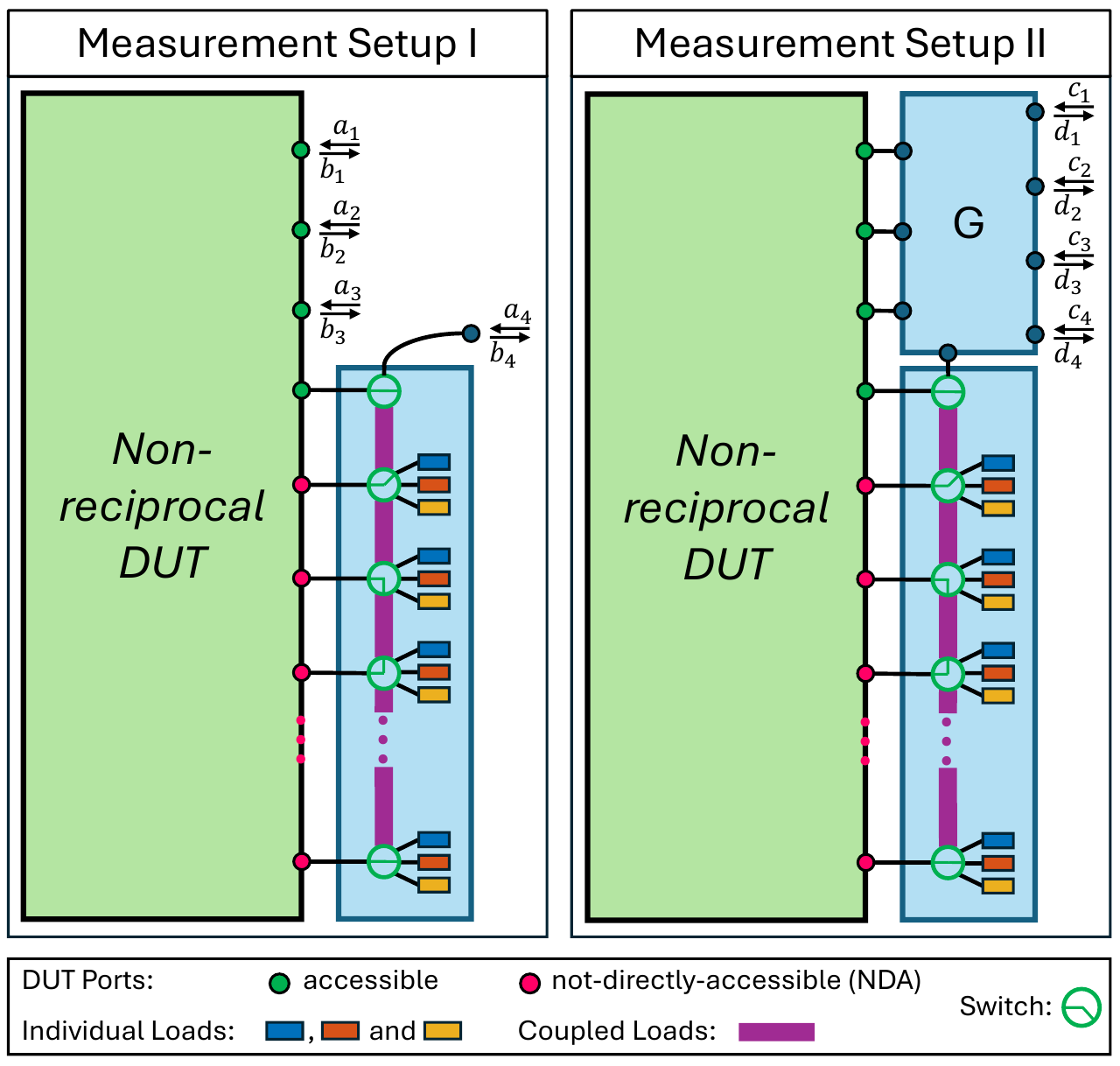}
    \caption{Overview of the two required ``Virtual VNA 3.1'' measurement setups. The goal is to determine the scattering matrix of a \textit{non-reciprocal} $N$-port DUT based on an $N_\mathrm{A}$-port coherent wavefront generator, an $N_\mathrm{A}$-port \textit{non}-coherent detector, and a specific $N_\mathrm{S}$-port tunable load network, where $1<N_\mathrm{A}<N$. The DUT ports are partitioned into $N_\mathrm{A}$ accessible ports (green) via which waves can be injected and received, and $N_\mathrm{S}=N-N_\mathrm{A}$ not-directly-accessible (NDA) ports (magenta) via which waves cannot be injected or received. The tunable load network  can terminate each NDA port with three distinct individual loads (blue, red, yellow), or connect pairs of neighboring NDA ports via a two-port load network (2PLN, purple); the last accessible and first NDA ports can also be connected by a 2PLN (purple). In Measurement Setup I (left panel), the coherent wavefront $\mathbf{a} = [a_1, a_2, a_3, a_4]$ is directly injected into the DUT's accessible ports, and the corresponding outgoing intensities $|\mathbf{b}|^2 = [|b_1|^2, |b_2|^2, |b_3|^2, |b_4|^2]$ are recorded. In Measurement Setup II (right panel), the accessible DUT ports are connected to a known $2N_\mathrm{A}$-port system G such that the coherent wavefront $\mathbf{c} = [c_1, c_2, c_3, c_4]$ is  injected via the free ports of G, and the corresponding outgoing intensities $|\mathbf{d}|^2 = [|d_1|^2, |d_2|^2, |d_3|^2, |d_4|^2]$ are recorded.  }
    \label{Fig1}
\end{figure}

As illustrated in Fig.~\ref{Fig1}, the tunable load network allows us to terminate the $i$th NDA port with one of three arbitrary but distinct individual loads (characterized by their known reflection coefficients $r_i^\mathrm{A}$, $r_i^\mathrm{B}$ and $r_i^\mathrm{C}$), or with a  coupled load (characterized by its known scattering matrix $\mathbf{S}^{\mathrm{2PLN,}i}\in\mathbb{C}^{2\times 2}$) connecting the $i$th NDA port to one of the neighboring DUT ports. There is no requirement for the three individual loads to coincide with calibration standards or to be the same at each NDA port. The coupled loads may be reciprocal or non-reciprocal and can be mutually distinct, as long as their transmission coefficients do not vanish (or else they cannot serve their purpose as coupled loads). The  $j$th configuration of the tunable load network is characterized by a block-diagonal scattering matrix $\mathbf{S}_{\mathrm{L},j}\in\mathbb{C}^{N_\mathrm{S}\times N_\mathrm{S}}$. If only individual loads are used to terminate the DUT's NDA ports, $\mathbf{S}_{\mathrm{L},j}$ is diagonal with the $i$th diagonal entry being the reflection coefficient of the load on the $i$th NDA port. 

Our Measurement Setup I is illustrated in the left panel of Fig.~\ref{Fig1}. A coherent wavefront generator allows us to inject an $N_\mathrm{A}$-element wavefront $\mathbf{a}\in\mathbb{C}^{N_\mathrm{A}}$ via the $N_\mathrm{A}$ accessible ports, and a \textit{non}-coherent detector allows us to measure the intensities (but not the phases) of the resulting outgoing wavefront $\mathbf{b}\in\mathbb{C}^{N_\mathrm{A}}$. 
For the $j$th configuration of the tunable load network, $\mathbf{b}=\hat{\mathbf{S}}_j\mathbf{a}$, where  $ \hat{\mathbf{S}}_j \in \mathbb{C}^{N_\mathrm{A} \times N_\mathrm{A}}$ characterizes the connection of the DUT and the tunable load network seen in the left panel of Fig.~\ref{Fig1}~\cite{anderson_cascade_1966,ha1981solid,prod2024efficient}:
\begin{equation}
    \hat{\mathbf{S}}_j(\mathbf{S},\mathbf{S}_{\mathrm{L},j}) = \mathbf{S}_\mathcal{AA} + \mathbf{S}_\mathcal{AS} \left(  \mathbf{S}_{\mathrm{L},j}^{-1} -  \mathbf{S}_\mathcal{SS} \right)^{-1}   \mathbf{S}_\mathcal{SA}.
    \label{eq_2}
\end{equation}
We denote by $\mathbf{A}_\mathcal{BC}$ the block of the matrix $\mathbf{A}$ whose row [column] indices are in the set $\mathcal{B}$ [$\mathcal{C}$].

In addition, the ``Virtual VNA 3.1'' requires a Measurement Setup II (illustrated in the right panel in Fig.~\ref{Fig1}) to lift the row-wise phase ambiguities described earlier. For Measurement Setup II, we connect a $2N_\mathrm{A}$-port auxiliary system G (characterized by its known scattering matrix $\mathbf{S}^\mathrm{G}\in\mathbb{C}^{2N_\mathrm{A}\times2N_\mathrm{A}}$) to the DUT's $N_\mathrm{A}$ accessible ports and inject a wavefront  $\mathbf{c}\in\mathbb{C}^{N_\mathrm{A}}$ via the remaining $N_\mathrm{A}$ free ports of G. A \textit{non}-coherent detector allows us to measure the intensities (but not the phases) of the resulting outgoing wavefront $\mathbf{d}\in\mathbb{C}^{N_\mathrm{A}}$. For the $j$th configuration of the tunable load network, $\mathbf{d}=\mathring{\mathbf{S}}_j\mathbf{c}$, where $\mathring{\mathbf{S}}_j\in\mathbb{C}^{N_\mathrm{A}\times N_\mathrm{A}}$ characterizes the connection of the DUT, the tunable load network and G seen in the right panel of Fig.~\ref{Fig1}:
\begin{equation}        \mathring{\mathbf{S}}_j(\mathbf{S},\mathbf{S}_{\mathrm{L},j},\mathbf{S}^\mathrm{G}) = \mathbf{S}^\mathrm{G}_\mathcal{FF} + \mathbf{S}^\mathrm{G}_{\mathcal{F}\bar{\mathcal{A}}} \left(  \hat{\mathbf{S}}_j^{-1} -  \mathbf{S}^\mathrm{G}_{\bar{\mathcal{A}}\bar{\mathcal{A}}} \right)^{-1}   \mathbf{S}^\mathrm{G}_{\bar{\mathcal{A}}\mathcal{F}},
    \label{eq_3}
\end{equation}
where $\bar{\mathcal{A}}$ and $\mathcal{F}$ are the sets of indices associated with the connected and free ports of G, respectively. G can be chosen arbitrarily as long as it results in interferences of the (incoming and) outgoing wavefronts, meaning that G cannot be realized with a collection of $N_\mathrm{A}$ uncoupled transmission lines. The requirement to know $\mathbf{S}^\mathrm{G}$ is not fundamentally problematic despite G having $2N_\mathrm{A}$ ports because we can choose a reciprocal G that we can characterize with the ``Virtual VNA 2.0'' technique for reciprocal DUTs~\cite{del2024virtual2p0}.

The problem that we tackle in this paper consists in identifying a suitable sequence of measurements with Measurement Setups I and II involving different configurations of the tunable load network to retrieve $\mathbf{S}$ fully and without ambiguity despite the constraint to non-coherent detection.

\section{Methods}
\label{sec_methods}

Our overall strategy is to adapt the gradient-descent ``Virtual VNA 3.0'' technique described in Sec. IV.B in~\cite{del2025virtual3p0} to the constraint of non-coherent detection. In~\cite{del2025virtual3p0}, an $N_\mathrm{A}$-port VNA was available such that only  Measurement Setup I was used. First, $M_1$ measurements of $\hat{\mathbf{S}}_j$ corresponding to random configurations of individual loads were conducted to retrieve an estimate of the sought-after $\mathbf{S}$ up to row-wise and column-wise scaling ambiguities. Second, $M_2$ measurements involving the first coupled load (between the last accessible and the first NDA port) and random configurations of individual loads on the other NDA ports were conducted to lift the scaling ambiguity on the row and column associated with the first NDA port. Third, $M_2$ measurements with random configurations of individual loads except for a 2PLN connecting two NDA ports were conducted for each remaining NDA port in turn to lift the remaining scaling ambiguities.

In this section, we begin by describing in Sec.~\ref{subsec_PR} a general method for retrieving without any ambiguity the phases of the entries of $\hat{\mathbf{S}}_j$ by leveraging known interferences of the outgoing wavefronts. Then, we use this method in Sec.~\ref{V2NA3p1} to adapt the ``Virtual VNA 3.0'' gradient-descent technique from~\cite{del2025virtual3p0} to non-coherent detection.

\subsection{Ambiguity-Free Phase Retrieval}
\label{subsec_PR}

In this subsection, we tackle the problem of unambiguously retrieving the phases of the measurable scattering matrix $\hat{\mathbf{S}}_j$ corresponding to the termination of the DUT's NDA ports with the $j$th configuration of the tunable load network. The tunable load network remains in its $j$th configuration throughout this subsection.

\textit{Step A: Estimate $\hat{\mathbf{S}}_j$ up to row-wise phase ambiguities.} 
Based on Measurement Setup I, we inject $N_\mathrm{pilot}$ known wavefronts $\mathbf{a}$ and detect the corresponding outgoing intensities $|\mathbf{b}|^2$. The injected wavefronts can be chosen randomly, only ``one-hot'' wavefronts with all entries except one being zero should be avoided because they do not allow for interferences between signals injected via different ports. Given $N_\mathrm{pilot}$ pairs $\{\mathbf{a},|\mathbf{b}|^2\}$, we use gradient descent to estimate $\hat{\mathbf{S}}_j$. To that end, we treat all entries of $\hat{\mathbf{S}}_j$ as trainable complex-valued weights (recall that there is no symmetry due to the DUT's non-reciprocity) and then minimize using gradient descent a cost function that quantifies the difference between predicted and measured $|\mathbf{b}|^2$, averaged over all $N_\mathrm{pilot}$ realizations of $\mathbf{a}$. This yields an estimate of $\hat{\mathbf{S}}_j$ with row-wise phase ambiguities, as explained in the introduction.

\textit{Step B: Estimate $\mathring{\mathbf{S}}_j$ up to row-wise phase ambiguities.} We apply the same procedure as in Step A to Measurement Setup II, yielding an estimate of $\mathring{\mathbf{S}}_j$ with row-wise phase ambiguities.

\textit{Step C: Lift row-wise phase ambiguities for $\hat{\mathbf{S}}_j$.} We now treat the $N_\mathrm{A}$ unknown phase factors, one for each of the $N_\mathrm{A}$ rows of $\hat{\mathbf{S}}_j$, as trainable weights. Using~(\ref{eq_3}), we obtain a prediction $\mathring{\mathbf{S}}_j^\mathrm{PRED}$ of $\mathring{\mathbf{S}}_j$ that depends on the phase factors that we seek to determine, our estimate $\hat{\mathbf{S}}_j^\mathrm{EST}$ of $\hat{\mathbf{S}}_j$ from Step A, and our knowledge of $\mathbf{S}^\mathrm{G}$. Recall that we retrieved an estimate $\mathring{\mathbf{S}}_j^\mathrm{EST}$ of $\mathring{\mathbf{S}}_j$ in Step B that suffers from row-wise phase ambiguities. We hence define a cost function $\mathcal{C}$ that quantifies the difference between $\mathring{\mathbf{S}}_j^\mathrm{PRED}$ and $\mathring{\mathbf{S}}_j^\mathrm{EST}$ while being insensitive to row-wise phase factors:
\begin{equation}
\label{eq_4}
\mathcal{C}\;=\;
\sum_{r=1}^{N_\mathrm{A}}
\Bigl(
\bigl\|\mathring{\mathbf{s}}_{j,r}^\mathrm{EST}\bigr\|^2
\;+\;
\bigl\|\mathring{\mathbf{s}}_{j,r}^\mathrm{PRED}\bigr\|^2
\;-\;
2\,\Bigl|
\mathring{\mathbf{s}}_{j,r}^\mathrm{EST}
\cdot
\overline{\mathring{\mathbf{s}}_{j,r}^\mathrm{PRED}}
\Bigr|
\Bigr),
\end{equation}
where \(\mathring{\mathbf{s}}_{j,r}^\mathrm{EST}\) and \(\mathring{\mathbf{s}}_{j,r}^\mathrm{PRED}\) denote the \(r\)th rows of \(\mathring{\mathbf{S}}_j^\mathrm{EST}\) and 
\(\mathring{\mathbf{S}}_j^\mathrm{PRED}\), respectively. 
The first two terms in the sum are squared Euclidean norms of vectors whereas the third term in the sum involves the absolute value of the inner product of two vectors. Using gradient descent, we retrieve the sought-after set of phase factors that is expected to minimize this cost function. 

The three steps described in this subsection allow us to estimate $\hat{\mathbf{S}}_j$ free of any ambiguity (i.e., not even a global phase ambiguity remains). The secondary measurement setup involving the known system G plays a key role because it creates known interferences of the (incoming and) outgoing wavefronts that allow us to lift the row-wise phase ambiguities. The described procedure rigorously accounts for multiple scattering without relying on an assumption of forward-only scattering.

\subsection{Virtual VNA 3.1}
\label{V2NA3p1}

\begin{figure}
    \centering
    \includegraphics[width=\columnwidth]{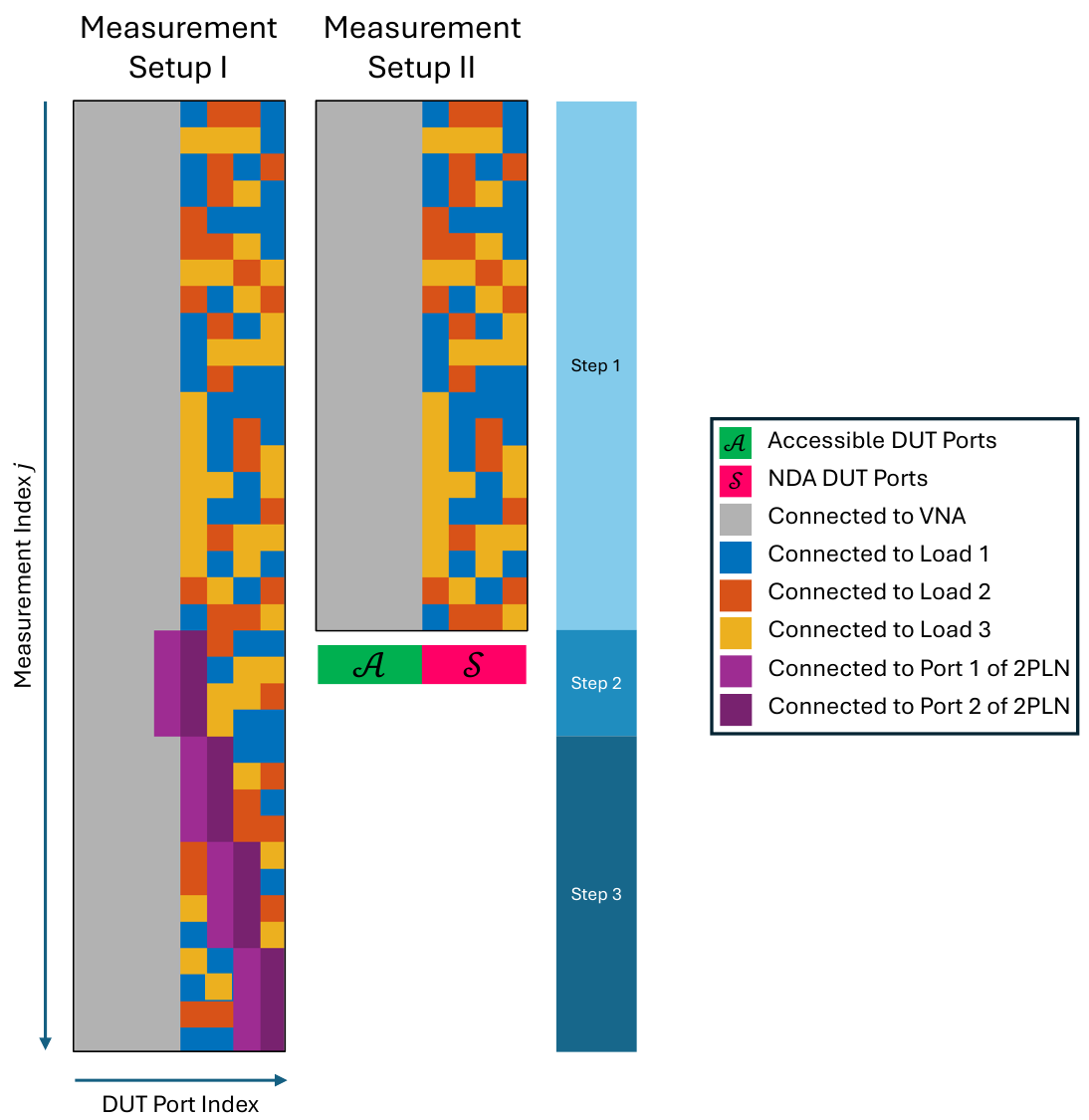}
    \caption{Measurement protocol based on the two measurement setups shown in Fig.~\ref{Fig1}. In Step 1, the same configurations of the tunable load network (only involving individual loads) are measured in both measurement setups. In Steps 2 and 3, the configurations of the tunable load network involve a coupled load and are only measured in Measurement Setup I.}
    \label{Fig2}
\end{figure}

Thanks to the procedure outlined in the previous subsection, we can retrieve $\hat{\mathbf{S}}_j$ without any ambiguity despite the constraint to non-coherent detection. 
Based on the procedure from the previous subsection, we hence recover exactly the same information that was directly provided by the VNA in~\cite{del2025virtual3p0}. The absence of global phase ambiguities facilitates the meaningful evaluation of changes of the measurable scattering matrix.\footnote{Our previous work~\cite{del2024virtual,del2024virtual2p0} on characterizing \textit{reciprocal} DUTs with non-coherent detection did not require Measurement Setup II. Consequently, each retrieved $\hat{\mathbf{S}}_j$ had a different random global phase offset. It was necessary to align the global phase offsets to meaningfully evaluate changes of the measurable scattering matrix.} Consequently, we can directly implement Step 1 analogous to~\cite{del2025virtual3p0}.

Specifically, in Step 1, we define a series of $M_1$ random configurations of the tunable load network involving random choices of the individual loads. For each configuration, we retrieve the corresponding $\hat{\mathbf{S}}_j$ using the procedure outline in Sec.~\ref{subsec_PR} which involves measurements in Measurement Setups I and II, as seen in Fig.~\ref{Fig2}. In each case, we check whether the minimized cost function $\mathcal{C}$ is below a stringent threshold; if not, we discard the measurement. 
Given the retained pairs $\{\hat{\mathbf{S}}_j,\mathbf{S}_{\mathrm{L},j}\}$, we retrieve an estimate of $\mathbf{S}$ up to row-wise and column-wise scaling ambiguities using the same gradient-descent procedure as in Step 1 in Sec.~IV.B in~\cite{del2025virtual3p0}.

To lift these scaling ambiguities, we use coupled loads as in~\cite{del2025virtual3p0}. In principle, we could again use both Measurement Setups I and II to obtain ambiguity-free estimates of the resulting measurable scattering matrices $\hat{\mathbf{S}}_j$. However, it is simpler and sufficient to merely measure the magnitudes of $\hat{\mathbf{S}}_j$ in Measurement Setup I. Hence, we do not use Measurement Setup II in Steps 2 and 3; instead, we adapt the corresponding cost function to compare only magnitudes of the predicted and measured $\hat{\mathbf{S}}_j$ (whereas Steps 2 and 3 in~\cite{del2025virtual3p0} compared both magnitudes and phases). 

Specifically, in Step 2, we connect the last accessible port and the first NDA port via a coupled load. For $M_2$ random configurations of individual loads terminating the other NDA ports, we measure the magnitudes of the corresponding $\hat{\mathbf{S}}_j$ in Measurement Setup I. We treat the scaling factor that lifts the row-wise and column-wise ambiguities associated with the first NDA port as a trainable variable. We evaluate the magnitudes of the entries of the difference between prediction and measurement of $|\hat{\mathbf{S}}_{j}|$ (rather than of $\hat{\mathbf{S}}_{j}$ as in~\cite{del2025virtual3p0}), and take the average thereof over all entries and the $M_2$ measurements. This quantity is our cost. We seek via gradient descent a value of our trainable variable that minimizes this cost. The resulting value of our trainable variable is our estimate of the sought-after scaling factor to lift the row-wise and column-wise scaling ambiguities associated with the first NDA port.

Then, in Step 3, we repeat similar procedures for each of the remaining NDA ports in turn. For the $i$th NDA port ($i>1$), we connect a coupled load to the $(i-1)$th and $i$th NDA ports. For $M_2$ random configurations of individual loads terminating the other NDA ports, we measure the magnitudes of the corresponding $\hat{\mathbf{S}}_j$ in Measurement Setup I. Then, using gradient descent, we lift the row-wise and column-wise scaling ambiguities of $\mathbf{S}$ associated with the $i$th NDA port. 

A visual summary of the complete ``Virtual VNA 3.1'' measurement protocol is provided in Fig.~\ref{Fig2} for the case of $N_\mathrm{A}=N_\mathrm{S}=4$ that we consider in our experiments in the next section.

\section{Experimental Validation}
\label{sec_ExpVal}

In this section, we validate the presented ``Virtual VNA 3.1'' technique based on experimental measurements, considering an eight-port non-reciprocal circuit with four accessible ports and four NDA ports as our DUT. We leverage the linearity of the wave equation to synthetically generate $N_\mathrm{pilot}=500$ pairs $\{\mathbf{a},|\mathbf{b}|^2\}$ or $\{\mathbf{c},|\mathbf{d}|^2\}$ based on VNA-based measurements of $\hat{\mathbf{S}}_j$ or $\mathring{\mathbf{S}}_j$, respectively.

\subsection{Experimental Setup}

\begin{figure}
    \centering
    \includegraphics[width=\columnwidth]{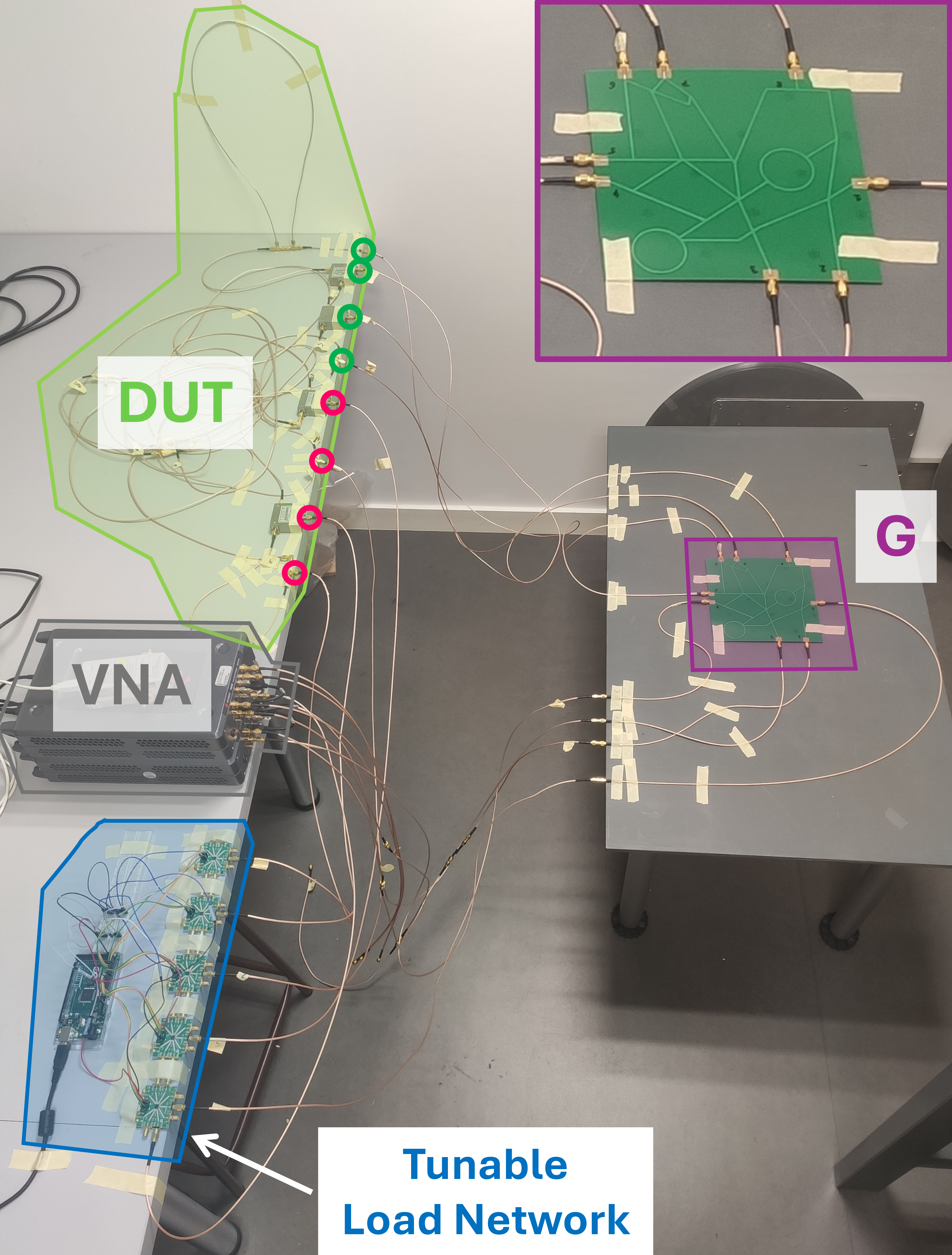}
    \caption{Photographic image of Measurement Setup II comprising an eight-port \textit{non-reciprocal} DUT (a complex transmission-line network involving four circulators), a four-port VNA, a tunable load network with four ``virtual VNA ports'' and the eight-port system G. The DUT's eight ports are partitioned into four accessible ports (green) and four NDA ports (magenta).
    The tunable load network comprises four switches (each connected to one of the DUT's four NDA ports) as well as an additional switch connected to the DUT's last accessible port, in line with the schematic in Fig.~\ref{Fig1}. Cables are treated as part of the VNA, the tunable load network or G.  We synthetically evaluate pairs of input and output wavefronts leveraging the linearity of the wave equation based on VNA-based measurements of four-port scattering matrices. The VNA has four additional ports which are only used to measure the ground-truth DUT scattering matrix to validate our results. }
    \label{Fig3}
\end{figure}

\begin{figure}
    \centering
    \includegraphics[width=\columnwidth]{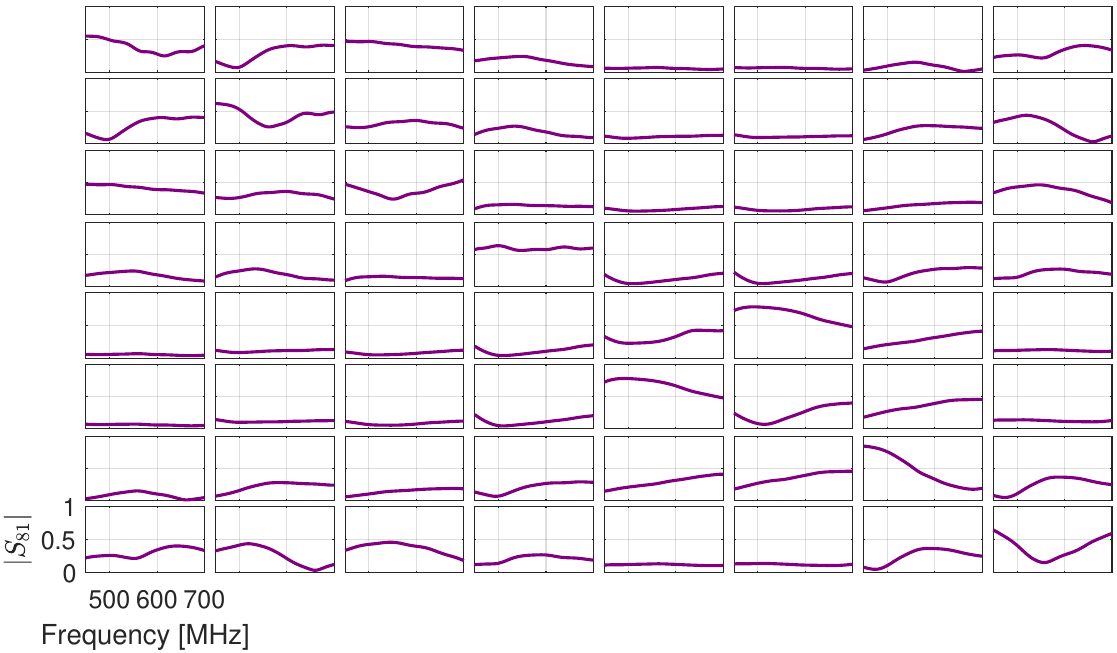}
    \caption{Measured magnitudes of the $8 \times 8$ scattering coefficient spectra for the reciprocal system G (depicted in the inset in Fig.~\ref{Fig3}) used in Measurement Setup II.}
    \label{Fig4}
\end{figure}

Our experimental setup is an extension of the setup used to validate the ``Virtual VNA 3.0'' technique in~\cite{del2025virtual3p0}. In particular, Measurement Setup I is analogous to the setup seen in Fig.~3 of~\cite{del2025virtual3p0} such that we do not reproduce it here. The utilized four-port VNA and the tunable load network offering four additional ``virtual'' VNA ports are exactly the same as in~\cite{del2025virtual3p0}. Moreover, the eight-port DUT based on a complex transmission-line network involving four circulators is almost the same as in~\cite{del2025virtual3p0}.\footnote{Our DUT can also be described as a non-reciprocal cable-network metamaterial or a non-reciprocal quantum-graph analogue~\cite{sol2024covert}.} We hence focus in this subsection on the  description  of Measurement Setup II. As seen in Fig.~\ref{Fig1}, Measurement Setup II involves an additional $2N_\mathrm{A}$-port system referred to as G that is connected to the DUT's accessible ports. As highlighted in purple and shown in the inset in Fig.~\ref{Fig3}, our realization of G involves a complex, reciprocal eight-port transmission line network printed on a circuit board. Its measured scattering characteristics are displayed in Fig.~\ref{Fig4}. It is apparent that this realization of G meets the requirement for creating interferences between signals injected via different ports. Moreover, it is clear that a forward-only scattering approximation would be unreasonable here, underlining the importance of our rigorous multi-port network formulation that properly accounts for multiple scattering between the DUT and G.

\subsection{Measurement Procedure}

For $N_\mathrm{S}=4$, $3^4=81$ distinct  configurations of the tunable load network exist that only involve individual loads. We hence choose $M_1=M_2=81$, using each of the 81 configurations exactly once. Because we manually switch between Measurement Setups I and II, we first perform all required measurements with the former and then all required measurements with the latter (see Fig.~\ref{Fig2} for an overview of the measurement procedure). Our measurement plane is at the DUT ports. Cables are treated as part of the VNA, the tunable load network or G. Before the main measurements, we characterize the tunable load network and G. We also measure the ground truth of $\mathbf{S}$ for validation purposes.

\subsection{Results}

\begin{figure*}
    \centering
    \includegraphics[width=\textwidth]{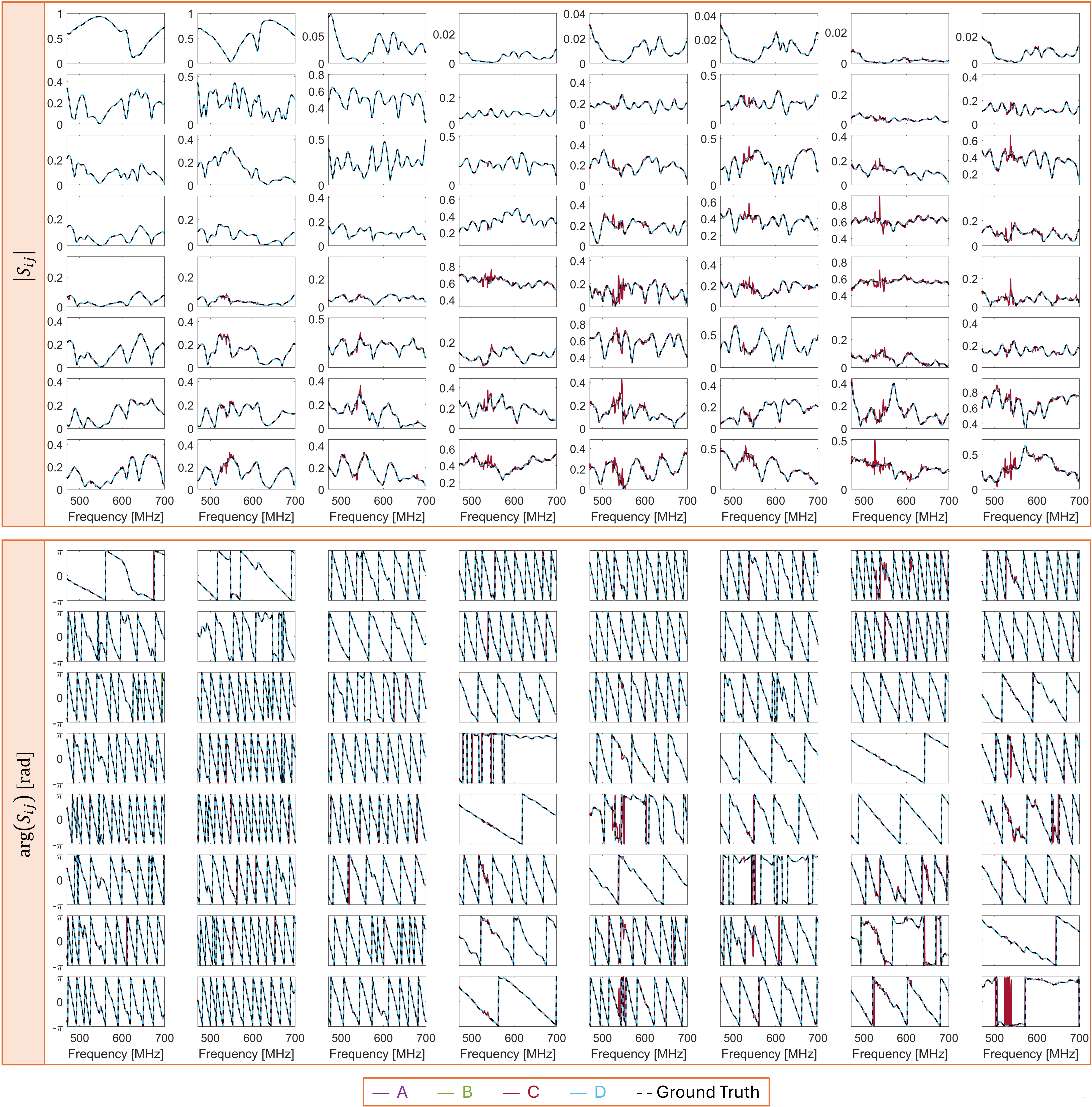}
    \caption{Comparison of four estimates for the \textit{non-reciprocal} DUT's $8\times8$ scattering spectra in terms of magnitudes (top panel) and phases (bottom panel). A [purple]: Virtual VNA 3.0 gradient-descent technique from~\cite{del2025virtual3p0}; B [green]: like A but using an intensity-only cost function in Steps 2 and 3; C [red]: Virtual VNA 3.1 for data synthesized based on measurements of $\hat{\mathbf{S}}_j$ or $\mathring{\mathbf{S}}_j$; D [light blue]: Virtual VNA 3.1 for data synthesized based on measurements of $\mathbf{S}$, $r_i^\mathrm{A}$, $r_i^\mathrm{B}$, $r_i^\mathrm{C}$, $\mathbf{S}^{\mathrm{2PLN},i}$ and $\mathbf{S}^\mathrm{G}$. The ground truth (black dashed line) is superimposed on top in each subplot. The purple (A) and green (B) lines are barely visible because they are almost perfectly covered by the light blue (D) line, which in turn agrees perfectly with the black-dashed line (groud truth); Approaches A, B and D are hence all three in very good agreement with the ground truth.}
    \label{Fig5}
\end{figure*}

\begin{figure*}
    \centering
    \includegraphics[width=\textwidth]{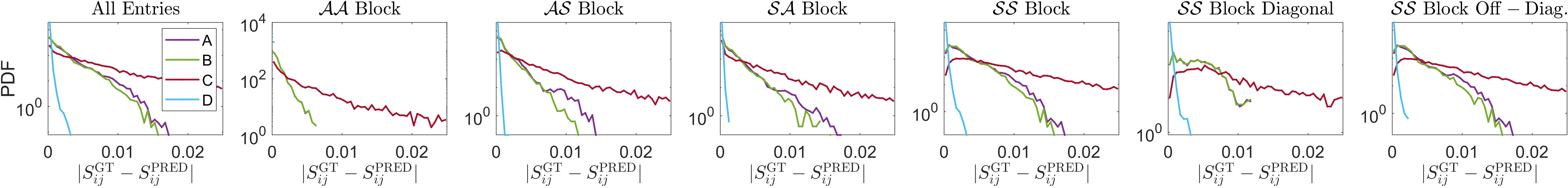}
    \caption{Probability density function (PDF) of the absolute error of the  scattering coefficients reconstructed with four approaches: A [purple]: Virtual VNA 3.0 gradient-descent technique from~\cite{del2025virtual3p0}; B [green]: like A but using an intensity-only cost function in Steps 2 and 3; C [red]: Virtual VNA 3.1 for data synthesized based on measurements of $\hat{\mathbf{S}}_j$ or $\mathring{\mathbf{S}}_j$; D [light blue]: Virtual VNA 3.1 for data synthesized based on measurements of $\mathbf{S}$, $r_i^\mathrm{A}$, $r_i^\mathrm{B}$, $r_i^\mathrm{C}$, $\mathbf{S}^{\mathrm{2PLN},i}$. The PDFs are evaluated across all 401 frequency points and additionally across all scattering coefficients or a subset thereof  (see subplot titles). }
    \label{Fig6}
\end{figure*}

To systematically evaluate the proposed ``Virtual VNA 3.1'' technique and to understand the influence of relevant factors on the achieved accuracy, we compare estimates of $\mathbf{S}$ obtained with the following four approaches:
\begin{enumerate}[label=\Alph*)]
    \item Virtual VNA 3.0 gradient-descent technique from~\cite{del2025virtual3p0}.
    \item Virtual VNA 3.0 gradient-descent technique from~\cite{del2025virtual3p0} except for using an intensity-only cost function in Steps 2 and 3.
    \item Virtual VNA 3.1 with synthetically generated realizations of $\{\mathbf{a},|\mathbf{b}|^2\}$ and $\{\mathbf{c},|\mathbf{d}|^2\}$ based on VNA-based experimental measurements of $\hat{\mathbf{S}}_j$ or $\mathring{\mathbf{S}}_j$, respectively.
    \item Virtual VNA 3.1 with synthetically generated realizations of $\{\mathbf{a},|\mathbf{b}|^2\}$ and $\{\mathbf{c},|\mathbf{d}|^2\}$ based on synthetically generated realizations of $\hat{\mathbf{S}}_j$ or $\mathring{\mathbf{S}}_j$, respectively, based on~(\ref{eq_2}) and~(\ref{eq_3}) in combination with VNA-based experimental measurements of $\mathbf{S}$, $r_i^\mathrm{A}$, $r_i^\mathrm{B}$, $r_i^\mathrm{C}$, $\mathbf{S}^{\mathrm{2PLN},i}$ and $\mathbf{S}^\mathrm{G}$.
\end{enumerate}
Approach A is the ``Virtual VNA 3.0'' benchmark for cases in which an $N_\mathrm{A}$-port VNA is available. The comparison between Approaches A and B informs us about possible accuracy penalties resulting from using an intensity-only cost function in Steps 2 and 3, as done in the proposed ``Virtual VNA 3.1'' instead of retrieving the complex-valued measurable scattering matrices in Steps 2 and 3. Approaches C and D are both based on the proposed ``Virtual VNA 3.1'' technique and are both grounded in experimental measurements; they differ regarding the use of experimentally measured versus synthetically generated realizations of $\hat{\mathbf{S}}_j$ and $\mathring{\mathbf{S}}_j$. Thereby, the comparison of Approaches C and D informs us about the vulnerability to experimental imperfections, resulting chiefly from the manual switching between Measurement Setups I and II in our case.

We begin by visually inspecting the 64 estimated scattering coefficient spectra in terms of their magnitudes and phases in Fig.~\ref{Fig5}. We superpose the four lines for Approaches A-D in Fig.~\ref{Fig5}, with the ground truth as black dashed line on top. Upon visual inspection, approaches A, B and D very closely resemble the ground truth. For Approach A, this is expected because it was already validated in~\cite{del2025virtual3p0} where it yielded very good accuracy. For Approach B, this means that using an intensity-only cost function in Steps 2 and 3 is indeed not problematic. For Approach D, this means that the proposed ``Virtual VNA 3.1'' technique works very well under ideal conditions. Meanwhile, for Approach C, we observe some deviations. The curves from Approach C generally follow the ground truth and are flawless in some frequency intervals but display noisy deviations in other frequency intervals. The noisy deviations are particularly strong in the vicinity of 530~MHz. It is not clear why the strength of these noisy deviations varies so significantly across the considered frequency range. Thanks to the comparison with Approach D, it is clear that these noisy deviations must originate from experimental inaccuracies that could arise, for instance, in manually switching between the two measurement setups. More specifically, we can pinpoint Step C from Sec.~\ref{subsec_PR} as the Achilles heel of the Virtual VNA 3.1. 

To examine the estimation accuracies more carefully, we plot in Fig.~\ref{Fig6} the probability density functions (PDFs) of the absolute error of the estimated scattering coefficients. The curves associated with Approach D are seen to drop very sharply, the curves associated with Approaches A and B closely resemble each other, and the curves associated with Approach C decay notably more slowly. The latter makes sense given the frequency intervals with noisy deviations observed in Fig.~\ref{Fig5}.

\begin{table}[b]
\centering
\caption{Performance of the four approaches (A, B, C, D) in terms of the $\zeta$ metric for different groups of scattering coefficients.}
\begin{tabular}{ |p{2.8cm}||>{\centering\arraybackslash}p{0.8cm}|>{\centering\arraybackslash}p{0.8cm}|>{\centering\arraybackslash}p{0.8cm}|>{\centering\arraybackslash}p{0.8cm}| }
 \hline
  & {A} & {B} & {C} & {D} \\
 \hline\hline
 {All Entries}                       & 38.5~dB    & 39.4~dB & 23.3~dB & 64.8~dB \\
 {\(\mathcal{AA}\) Block}            & 47.6~dB    & 47.6~dB & 30.0~dB & 87.1~dB \\
 \(\mathcal{AS}\) Block              & 38.4~dB    & 40.5~dB & 24.5~dB & 67.4~dB \\
 \(\mathcal{SA}\) Block              & 38.8~dB    & 39.9~dB & 24.6~dB & 68.1~dB \\
 \(\mathcal{SS}\) Block              & 37.3~dB    & 38.3~dB & 22.2~dB & 63.5~dB \\
 \(\mathcal{SS}\) Block Diagonal     & 33.3~dB    & 33.3~dB & 16.3~dB & 56.8~dB \\
 \(\mathcal{SS}\) Block Off-Diagonal & 37.7~dB    & 38.9~dB & 23.2~dB & 64.9~dB \\
 \hline
\end{tabular}
\label{table_zeta}
\end{table}

To quantify the accuracy of an estimate of $\mathbf{S}$ with a single metric, we evaluate
\begin{equation}
    \zeta = \left\langle\frac{\mathrm{SD}\left[S_{ij}^\mathrm{GT}(f)\right]}{\mathrm{SD}\left[S_{ij}^\mathrm{GT}(f) - S_{ij}^\mathrm{PRED}(f)\right]}\right\rangle_{i,j,f},
    \label{eq_zeta}
\end{equation}
where $\mathrm{SD}$ denotes the standard deviation, and the superscripts GT and PRED denote ground truth and prediction, respectively. Higher values of $\zeta$ indicate a higher accuracy because the definition in (\ref{eq_zeta}) resembles a signal-to-noise ratio, except that the mismatch between prediction and ground truth plays the role of the ``noise''.
A general observation is that for any approach, $\zeta$ is highest for the $\mathcal{AA}$ block and lowest for the diagonal entries of the $\mathcal{SS}$ block. The values of  $\zeta$ for Approaches A and B are very similar among each other and to those from~\cite{del2025virtual3p0}. Interestingly, the values of  $\zeta$ are sometimes even slightly higher for Approach B despite the latter's intensity-only cost function in Steps 2 and 3. For Approach D, the values of $\zeta$  are extraordinarily high because no experimental inaccuracies arising from manually connections can impact Approach D. For Approach C, the values of $\zeta$ are notably lower, as expected upon inspection of Fig.~\ref{Fig5}.

Before closing, we note that the noisy deviations in Approach C are indeed ``noisy'', i.e., almost uncorrelated for nearby frequency points. Therefore, they can be detected easily (even without comparison to the ground truth), and maybe even fixed in future work (see outlook in Sec.~\ref{sec_conclusion} below).

\section{Conclusion}
\label{sec_conclusion}

To summarize, we have introduced the ``Virtual VNA 3.1'' technique for unambiguously estimating a \textit{non-reciprocal} DUT's full scattering matrix by only injecting or receiving waves through a fixed subset of ``accessible'' DUT ports under a constraint to \textit{non}-coherent detection. On the one hand, our technique requires the termination of the DUT's ``not-directly-accessible'' ports with a specific tunable load network, akin to previous ``Virtual VNA'' protocols~\cite{del2024virtual2p0,del2025virtual3p0,kitvna}. On the other hand, our technique requires an additional modified measurement setup in which a known system is connected to the DUT's accessible ports to create known interferences of the outgoing waves. This second measurement setup is necessary to resolve row-wise phase ambiguities arising under a constraint to \textit{non}-coherent detection when the DUT is \textit{non-reciprocal}.

The developed method for \textit{non-reciprocal} DUTs can also be applied to reciprocal DUTs; however, if the DUT is known to be reciprocal, this a priori knowledge should be leveraged to constrain the estimation problem. For reciprocal DUTs, it is easier to use the ``Virtual VNA 2.0''~\cite{del2024virtual2p0} protocol for non-coherent detection because it does not require Measurement Setup II.

Considering an eight-port complex non-reciprocal circuit, we validated the developed ``Virtual VNA 3.1'' technique based on experimental measurements. Our experiments relied on manual switching between the two measurement setups, which revealed a vulnerability of the proposed technique to minute experimental inaccuracies that can result in noisy deviations of the estimated scattering coefficients from the ground truth.

Looking forward, \textit{on the hardware side}, we believe that the accuracy can be notably improved with a printed circuit board implementing the tunable load network, the known system to be connected to the DUT's accessible ports in the second measurement setup, and the switching between the two measurement setups.  Meanwhile, \textit{on the software side}, we expect that spectral correlations can be leveraged to efficiently detect and remove the observed noisy deviations from the ground truth. Indeed, in the present paper, we treated each frequency point independently. By leveraging the knowledge that the dispersion of the scattering coefficients must be a smooth function, we believe that the noisy deviations can be removed with additional post-processing steps. Such techniques would incidentally benefit the ``Virtual VNA'' family~\cite{del2024virtual,del2024virtual2p0,del2025virtual3p0,kitvna} more broadly, not being restricted to \textit{non-reciprocal} DUTs under a constraint of \textit{non}-coherent detection. So far, only~\cite{kitvna} made limited use of spectral correlations to improve the robustness of lifting sign ambiguities in the case of reciprocal DUTs.

\bibliographystyle{IEEEtran}

\begin{thebibliography}{10}
\providecommand{\url}[1]{#1}
\csname url@samestyle\endcsname
\providecommand{\newblock}{\relax}
\providecommand{\bibinfo}[2]{#2}
\providecommand{\BIBentrySTDinterwordspacing}{\spaceskip=0pt\relax}
\providecommand{\BIBentryALTinterwordstretchfactor}{4}
\providecommand{\BIBentryALTinterwordspacing}{\spaceskip=\fontdimen2\font plus
\BIBentryALTinterwordstretchfactor\fontdimen3\font minus \fontdimen4\font\relax}
\providecommand{\BIBforeignlanguage}[2]{{%
\expandafter\ifx\csname l@#1\endcsname\relax
\typeout{** WARNING: IEEEtran.bst: No hyphenation pattern has been}%
\typeout{** loaded for the language `#1'. Using the pattern for}%
\typeout{** the default language instead.}%
\else
\language=\csname l@#1\endcsname
\fi
#2}}
\providecommand{\BIBdecl}{\relax}
\BIBdecl

\bibitem{tippet1982rigorous}
J.~C. Tippet and R.~A. Speciale, ``A rigorous technique for measuring the scattering matrix of a multiport device with a 2-port network analyzer,'' \emph{IEEE Trans. Microw. Theory Tech.}, vol.~30, no.~5, pp. 661--666, 1982.

\bibitem{ruttan2008multiport}
T.~G. Ruttan, B.~Grossman, A.~Ferrero, V.~Teppati, and J.~Martens, ``Multiport {VNA} measurement,'' \emph{IEEE Microw. Mag.}, vol.~9, no.~3, pp. 56--69, 2008.

\bibitem{2023paper}
Y.-C. Chien, H.-W. Liu, and T.-L. Wu, ``Reducing the number of measurements of a multiport circuit using covering designs and {Turán} systems,'' \emph{IEEE Microw. Wirel. Technol. Lett.}, vol.~33, no.~2, pp. 115--117, 2023.

\bibitem{del2024virtual}
P.~del Hougne, ``{Virtual VNA}: Minimal-ambiguity scattering matrix estimation with a fixed set of ``virtual'' load-tunable ports,'' \emph{IEEE Trans. Instrum. Meas.}, 2025.

\bibitem{del2024virtual2p0}
------, ``Virtual {VNA 2.0}: Ambiguity-free scattering matrix estimation by terminating not-directly-accessible ports with tunable and coupled loads,'' \emph{IEEE Trans. Antennas Propag.}, 2025.

\bibitem{del2025virtual3p0}
------, ``Virtual {VNA 3.0}: Unambiguous scattering matrix estimation for non-reciprocal systems by leveraging tunable and coupled loads,'' \emph{arXiv:2503.07239}, 2025.

\bibitem{kitvna}
J.~Tapie and P.~del Hougne, ``Scalable multiport antenna array characterization with {PCB}-realized tunable load network providing additional “virtual” {VNA} ports,'' \emph{arXiv:2503.16256}, 2025.

\bibitem{garbacz1964determination}
R.~Garbacz, ``Determination of antenna parameters by scattering cross-section measurements,'' in \emph{Proc. IEE}, vol. 111, no.~10.\hskip 1em plus 0.5em minus 0.4em\relax IET, 1964, pp. 1679--1686.

\bibitem{bauer1974embedding}
R.~F. Bauer and P.~Penfield, ``De-embedding and unterminating,'' \emph{IEEE Trans. Microw. Theory Techn.}, vol.~22, no.~3, pp. 282--288, 1974.

\bibitem{mayhan1994technique}
J.~T. Mayhan, A.~R. Dion, and A.~J. Simmons, ``A technique for measuring antenna drive port impedance using backscatter data,'' \emph{IEEE Trans. Antennas Propag.}, vol.~42, no.~4, pp. 526--533, 1994.

\bibitem{davidovitz1995reconstruction}
M.~Davidovitz, ``Reconstruction of the {S}-matrix for a 3-port using measurements at only two ports,'' \emph{IEEE Microw. Guid. Wave Lett.}, vol.~5, no.~10, pp. 349--350, 1995.

\bibitem{wiesbeck1998wide}
W.~Wiesbeck and E.~Heidrich, ``Wide-band multiport antenna characterization by polarimetric {RCS} measurements,'' \emph{IEEE Trans. Antennas Propag.}, vol.~46, no.~3, pp. 341--350, 1998.

\bibitem{lu2000port}
H.-C. Lu and T.-H. Chu, ``Port reduction methods for scattering matrix measurement of an n-port network,'' \emph{IEEE Trans. Microw. Theory Techn.}, vol.~48, no.~6, pp. 959--968, 2000.

\bibitem{lu2003multiport}
------, ``Multiport scattering matrix measurement using a reduced-port network analyzer,'' \emph{IEEE Trans. Microw. Theory Techn.}, vol.~51, no.~5, pp. 1525--1533, 2003.

\bibitem{pfeiffer2005characterization}
U.~R. Pfeiffer and A.~Chandrasekhar, ``Characterization of flip-chip interconnects up to millimeter-wave frequencies based on a nondestructive in situ approach,'' \emph{IEEE Trans. Adv. Packag.}, vol.~28, no.~2, pp. 160--167, 2005.

\bibitem{pfeiffer2005recursive}
U.~R. Pfeiffer and C.~Schuster, ``A recursive un-termination method for nondestructive in situ {S}-parameter measurement of hermetically encapsulated packages,'' \emph{IEEE Trans. Microw. Theory Techn.}, vol.~53, no.~6, pp. 1845--1855, 2005.

\bibitem{pfeiffer2005equivalent}
U.~Pfeiffer and B.~Welch, ``Equivalent circuit model extraction of flip-chip ball interconnects based on direct probing techniques,'' \emph{IEEE Microw. Wirel. Compon. Lett.}, vol.~15, no.~9, pp. 594--596, 2005.

\bibitem{pursula2008backscattering}
P.~Pursula, D.~Sandstrom, and K.~Jaakkola, ``Backscattering-based measurement of reactive antenna input impedance,'' \emph{IEEE Trans. Antennas Propag.}, vol.~56, no.~2, pp. 469--474, 2008.

\bibitem{bories2010small}
S.~Bories, M.~Hachemi, K.~H. Khlifa, and C.~Delaveaud, ``Small antennas impedance and gain characterization using backscattering measurements,'' \emph{Proc. EuCAP}, 2010.

\bibitem{denicke2012application}
E.~Denicke, M.~Henning, H.~Rabe, and B.~Geck, ``The application of multiport theory for {MIMO RFID} backscatter channel measurements,'' \emph{Proc. EuMC}, pp. 522--525, 2012.

\bibitem{monsalve2013multiport}
B.~Monsalve, S.~Blanch, and J.~Romeu, ``Multiport small integrated antenna impedance matrix measurement by backscattering modulation,'' \emph{IEEE Trans. Antennas Propag.}, vol.~61, no.~4, pp. 2034--2042, 2013.

\bibitem{van2020verification}
A.~J. Van Den~Biggelaar, E.~Galesloot, A.~C. Franciscus, A.~B. Smolders, and U.~Johannsen, ``Verification of a contactless characterization method for millimeter-wave integrated antennas,'' \emph{IEEE Trans. Antennas Propag.}, vol.~68, no.~5, pp. 3358--3365, 2020.

\bibitem{sahin2021noncontact}
S.~Sahin, N.~K. Nahar, and K.~Sertel, ``Noncontact characterization of antenna parameters in {mmW} and {THz} bands,'' \emph{IEEE Trans. Terahertz Sci. Technol.}, vol.~12, no.~1, pp. 42--52, 2021.

\bibitem{buck2022measuring}
D.~Buck, K.~F. Warnick, R.~Maaskant, D.~B. Davidson, and D.~F. Kelley, ``Measuring array mutual impedances using embedded element patterns,'' \emph{IEEE Trans. Antennas Propag.}, vol.~71, no.~1, pp. 606--611, 2022.

\bibitem{kruglov2023contactless}
D.~Kruglov, P.~Krasov, O.~Iupikov, A.~Vilenskiy, M.~Ivashina, and R.~Maaskant, ``Contactless measurement of a {D}-band on-chip antenna using an integrated reflective load switch,'' \emph{IEEE Antennas Wirel. Propag. Lett.}, vol.~23, no.~3, pp. 1075--1079, 2023.

\bibitem{sol2024experimentally}
J.~Sol, H.~Prod’homme, L.~Le~Magoarou, and P.~del Hougne, ``Experimentally realized physical-model-based frugal wave control in metasurface-programmable complex media,'' \emph{Nat. Commun.}, vol.~15, no.~1, p. 2841, 2024.

\bibitem{sol2024optimal}
J.~Sol, L.~Le~Magoarou, and P.~del Hougne, ``Optimal blind focusing on perturbation-inducing targets in sub-unitary complex media,'' \emph{Laser Photonics Rev.}, p. 2400619, 2024.

\bibitem{shilinkov2024antenna}
I.~Shilinkov and R.~Maaskant, ``Antenna array measurements by a scalable backscatter modulation procedure,'' \emph{IEEE Antennas Wirel. Propag. Lett.}, vol.~23, no.~10, pp. 2989--2993, 2024.

\bibitem{del2025physics}
P.~del Hougne, ``A physics-compliant diagonal representation for wireless channels parametrized by beyond-diagonal reconfigurable intelligent surfaces,'' \emph{IEEE Trans. Wirel. Commun.}, 2025.

\bibitem{shechtman2015phase}
Y.~Shechtman, Y.~C. Eldar, O.~Cohen, H.~N. Chapman, J.~Miao, and M.~Segev, ``Phase retrieval with application to optical imaging: a contemporary overview,'' \emph{IEEE Signal Process. Mag.}, vol.~32, no.~3, pp. 87--109, 2015.

\bibitem{dong2023phase}
J.~Dong, L.~Valzania, A.~Maillard, T.-a. Pham, S.~Gigan, and M.~Unser, ``Phase retrieval: From computational imaging to machine learning: A tutorial,'' \emph{IEEE Signal Process. Mag.}, vol.~40, no.~1, pp. 45--57, 2023.

\bibitem{dremeau2015reference}
A.~Dr{\'e}meau, A.~Liutkus, D.~Martina, O.~Katz, C.~Sch{\"u}lke, F.~Krzakala, S.~Gigan, and L.~Daudet, ``Reference-less measurement of the transmission matrix of a highly scattering material using a {DMD} and phase retrieval techniques,'' \emph{Opt. Express}, vol.~23, no.~9, pp. 11\,898--11\,911, 2015.

\bibitem{metzler2017coherent}
C.~A. Metzler, M.~K. Sharma, S.~Nagesh, R.~G. Baraniuk, O.~Cossairt, and A.~Veeraraghavan, ``Coherent inverse scattering via transmission matrices: Efficient phase retrieval algorithms and a public dataset,'' \emph{Proc. ICCP}, pp. 1--16, 2017.

\bibitem{caramazza2019transmission}
P.~Caramazza, O.~Moran, R.~Murray-Smith, and D.~Faccio, ``Transmission of natural scene images through a multimode fibre,'' \emph{Nat. Commun.}, vol.~10, no.~1, p. 2029, 2019.

\bibitem{del2016intensity}
P.~del Hougne, B.~Rajaei, L.~Daudet, and G.~Lerosey, ``Intensity-only measurement of partially uncontrollable transmission matrix: demonstration with wave-field shaping in a microwave cavity,'' \emph{Opt. Express}, vol.~24, no.~16, pp. 18\,631--18\,641, 2016.

\bibitem{goel2023referenceless}
S.~Goel, C.~Conti, S.~Leedumrongwatthanakun, and M.~Malik, ``Referenceless characterization of complex media using physics-informed neural networks,'' \emph{Opt. Express}, vol.~31, no.~20, pp. 32\,824--32\,839, 2023.

\bibitem{anderson_cascade_1966}
B.~D.~O. Anderson and R.~W. Newcomb, ``\BIBforeignlanguage{en}{Cascade connection for time-invariant n-port networks},'' \emph{\BIBforeignlanguage{en}{Proc. Inst. Electr. Eng.}}, vol. 113, no.~6, pp. 970--974, Jun. 1966.

\bibitem{ha1981solid}
T.~T. Ha, \emph{{Solid-State Microwave Amplifier Design}}. Wiley-Interscience, 1981.

\bibitem{prod2024efficient}
H.~Prod'homme and P.~del Hougne, ``Updatable closed-form evaluation of arbitrarily complex multi-port network connections,'' \emph{arXiv:2412.17884}, 2024.

\bibitem{sol2024covert}
J.~Sol, M.~R{\"o}ntgen, and P.~del Hougne, ``Covert scattering control in metamaterials with non-locally encoded hidden symmetry,'' \emph{Adv. Mater.}, vol.~36, no.~11, p. 2303891, 2024.

\end{thebibliography}

\providecommand{\noopsort}[1]{}\providecommand{\singleletter}[1]{#1}%

\end{document}